\documentclass[aps,prc,twocolumn,tightenlines,superscriptaddress,nofootinbib,a4paper,showpacs]{revtex4-2}
\usepackage{dcolumn}
\usepackage{xcolor}
\usepackage{graphicx}
\usepackage{rotating}
\usepackage{natbib}
\usepackage{amsmath}
\usepackage{braket}

\begin{document}

\title{Updated systematics of intermediate-energy single-nucleon removal cross sections}

\author{J.~A.~Tostevin}
\affiliation{Department of Physics, Faculty of Engineering and
Physical Sciences, University of Surrey, Guildford, Surrey GU2 7XH,
United Kingdom}

\author{A.~Gade}
\affiliation{National Superconducting Cyclotron Laboratory, Michigan
State University, East Lansing, Michigan 48824, USA}
\affiliation{Department of Physics and Astronomy, Michigan State
University, East Lansing, Michigan 48824, USA}
\date{\today}

\begin{abstract}
The body of experimental measurements of intermediate-energy reactions that
remove a single nucleon from a secondary beam of neutron- or proton-rich
nuclei continues to grow. These data have been analysed consistently using
an approximate, eikonal-model treatment of the reaction dynamics combined
with appropriate shell-model descriptions of the projectile initial state,
the bound final states spectrum of the reaction residue and single-particle
removal strengths computed from their wave-function overlaps. The systematics
of the ratio $R_s$ of the measured inclusive cross-section to all bound final
states and the calculated cross-section to bound shell-model states -- in
different regions of the nuclear chart and involving both very weakly-bound
and strongly-bound valence nucleons -- is important in relating the empirically
deduced orbital occupancies to those from the best available shell-model
predictions. Importantly, several new higher-energy measurements, for which
the sudden-approximation aspect of the dynamical description is placed on an
even stronger footing, now supplement the previously-analysed measurements.
These additional data sets are discussed. Their $R_s$ values are shown to
conform to and reinforce the earlier-observed systematics, with no indication
that the approximately linear reduction in $R_s$ with increasing nucleon
separation energy is a consequence of a breakdown of the sudden approximation.
\end{abstract}

\maketitle

\section{Introduction}

Measurements and eikonal-model analyses of single-nucleon removal reactions
from rare-isotope beams at intermediate energy, energies in excess of 80
MeV/nucleon, have been invaluable in advancing our understanding of
single-particle degrees of freedom and the evolution of shell structure
with increasing neutron-proton number asymmetry \cite{GG}. This, in turn,
has helped advance the development of new shell-model effective interactions;
see for example Refs.\ \cite{Uts01,Pov12,Cau14,ekk}. A global feature of
this increasing body of precision removal-reaction data is the behaviour
of the ratio, denoted $R_s$, of the measured and eikonal-model absolute
inclusive cross sections to all bound final states of the residual nucleus.
This behaviour is seen in the totality of data from very different regions
of the nuclear chart and including nuclei with extreme $A:Z$ number ratios.
These involve nucleon removals from both very weakly-bound and well-bound
valence single-particle orbitals near their respective Fermi surfaces. These
deduced ratios, as a function of the separation energy asymmetry $\Delta S$
of the two nucleon species, defined below, shows a characteristic and
essentially linear behaviour; decreasing from values near unity for removals
of the most weakly-bound nucleons to values $\approx 0.3(1)$ for the most
strongly-bound nucleon cases.

A first compilation, noting this behaviour and detailing the model calculations,
was presented in Ref. \cite{Gad08}. That work included three data points with
energies between 60 and 70 MeV/nucleon and also incorporated the earlier analysis
\cite{Bro02} of high-energy data for the $^{12}$C($-n,-p$) and $^{16}$O($-n,-p$)
reactions, at 250, 1050 and 2100 MeV/nucleon. The latter showed consistency with
the analogous cross-section ratios deduced from high-energy electron-induced proton
knockout from these and other stable nuclei. This initial data compilation was
subsequently enhanced in Ref. \cite{TG}, with the addition of new data analyses,
the majority of which were from measurements in the $80-120$ MeV/nucleon range.
Here, we add additional information to these growing systematics. Importantly,
we include several additional data points derived from new measurements at
significantly higher-energies. These now include reactions at the positive and
negative extremes of $\Delta S$.

\section{Reaction Model}

The eikonal-model theoretical description of the nucleon removal reaction
dynamics used in the analyses of these collisions -- of a projectile of mass $A$
with a light target nucleus -- uses the sudden (fast collision) and eikonal
(forward scattering) approximations. These ingredients have been presented
and discussed in detail elsewhere; see for example Refs. \cite{Tos01,Han03}
and the references therein. The key dynamical ingredients that enter the model
cross-section calculations are the elastic S-matrices of the removed nucleon
and the mass $A$$-$1 reaction residue, expressed as a function of their
collision impact parameters. These S-matrices describe the degree of transmission
and absorption of these particles as they transit the complex (optical-model)
interactions with the target nucleus. These highly-absorptive optical
interactions, in particular that of the ion-ion (residue-target) system, dictate
that the single-nucleon removing events, that require the transmission
(survival) of the  mass $A$$-$1 residues, are entirely dominated by impact
parameters that involve only grazing contact of the nuclear surfaces and
that do not penetrate into the nuclear interior. Such interior collisions, at
smaller impact parameters, result in more complex fragmentation channels with
essentially complete absorption of the mass $A$$-$1 residues. As a result of
this surface dominance of the single-nucleon removal mechanisms, the
strong-interaction path lengths along which the nucleon and residual nucleus
transit their optical potentials with the light-target nucleus are of short-range
and the interaction time is correspondingly short.

\subsection{Sudden approximation}

This strongly-absorptive potential behaviour, the geometry of the optical
potentials, and the surface dominance of the removal reaction events are largely
unchanged in the calculations for all incident energies of interest, i.e. those
in excess of 80 MeV/nucleon. So, the collision or transit time is essentially
dictated by the energy (velocity) of the projectile beam. For example, at 100
MeV/nucleon, with $\gamma=1.1$, $v/c\approx 0.4$ and the estimated collision
time ${\cal T}_{\rm coll}$ is of order 8$\times d \times 10^{-24}$ s, where $d$
(in fm) is the typical strong-interaction path length. Given the light target nuclei
used, with root mean squared (rms) matter radii of 2.36 fm (Be) and 2.32 fm (C), 
these expected path lengths are of order 2--4 fm. Such collision times are faster than conventional
light-ion direct-reaction times and those typically associated with any significant
motion of valence nucleons near the nuclear surface (and the Fermi surface) in the
nuclear ground state. These timescales are the basis of the sudden (sometimes
referred to as the fast-adiabatic) approximation -- that the vector separation
between the removed nucleon and the residual nucleons in the projectile ground-state
can be treated as frozen for the short duration of the collision, ${\cal T}_{\rm coll}$.

The magnitude of the leading-order correction to this sudden (fast-adiabatic)
approximation in the case of the simpler elastic breakup mechanism was
considered in some detail by Johnson \cite{RCJ} and Summers {\em et al.}
\cite{SJ} for weakly-bound projectiles. The elastic breakup mechanism is an
important, although not the dominant removal mechanism in well-bound nucleon
removal cases, as is quantified below. Importantly however, these studies
also concluded, as was argued above, that the spatial localization of this
removal reaction mechanism arising from the strongly-absorptive residue-target
interaction, removes the major part of the non-adiabatic corrections -- the
maximum of the leading correction term lying at smaller impact parameters at
which the strong-absorption has reduced the S-matrix is essentially zero.
Such a formal analysis is very difficult in the case of the inelastic breakup
mechanism where the presence of a continuum of target excitations to unbound
configurations destroys the simpler, three-body-like nature of the elastic
breakup dynamics. Hence, we confine discussion to the short collision time that
underpins the sudden approximation, dictated by the energy of the projectiles.

Clearly, this collision time ${\cal T}_{\rm coll}$ becomes shorter as the
beam energy is raised. This reduction is a by a factor of $\approx$2/3 when
the beam energy exceeds 240 MeV/nucleon, For more highly-relativistic, 1.6
GeV/nucleon projectiles, the energy of several new precision measurements
at which $v/c\approx 0.92$, ${\cal T}_{\rm coll}$ is reduced by more than a
factor of 2. So, a comparison of the cross-sections and $R_s$ deduced from the
existing data (most at $\approx$100 MeV/nucleon) and higher-energy measurements over
such an extended energy range may be expected to reveal differences if there
is any significant breakdown of the accuracy of the sudden approximation in
the analyses of the lower-energy data points. This comparison motivates the
present study.

\subsection{Other reaction considerations}

The intermediate energy of the projectile beam is essential for
the sudden and eikonal dynamical approximations to be applicable. A
minimum beam energy is also of particular importance for the points with large
positive $\Delta S$, the cases where the removed nucleons are most strongly bound.
The kinematics in these large negative Q-value reaction cases naturally imposes
an upper-limit upon the longitudinal-momentum carried by the fast, forward-traveling
reaction residues. This was shown dramatically in Ref. \cite{Fla12}, where the
$^{14}$O($-n$) reaction, with $S_n= 23.2$ MeV, was performed at too low a
secondary beam energy of 53 MeV/nucleon. There, this kinematical cutoff was
sufficiently low that it intruded into and distorted the conventionally
Gaussian-like longitudinal momentum distribution of the $^{13}$O residues. An
expression for this maximal beam-direction residue momentum was given in Ref.
\cite{TG}. That analysis, like that of Ref. \cite{Fla12}, concluded that, even
for the maximum nucleon separation energies expected physically, this kinematic
cutoff has minimal effect for reactions with beam energies close-to and in excess
of 80 MeV/nucleon. So, this kinematics-cutoff effect has no implications for
the higher-energy data sets we discuss here.

The additional dynamical approximation made in the model calculations is that
the mass $A-1$ reaction residue is treated as a spectator and its internal
state, $\alpha$, is preserved in the collision. Use of the spectator approximation
is critical to the spectroscopic usefulness of the reaction, since the yield of
residues in a particular final state $\alpha$ then reflects the component
(parentage) of this configuration in the ground-state wave function of the
projectile. This assumption is not explicitly studied here and is assumed in
all calculations, regardless of the separation energy of the removed nucleon.
Nevertheless, since measured and calculated ion-ion inelastic cross sections
fall with increasing collision energy, it is expected that the spectator
approximation will also improve with the projectile energy.

It is important to remember that the measured and calculated removal cross
sections are highly inclusive with respect to the (unobserved) final
states of the target nucleus, being $^9$Be and $^{12}$C in all cases considered.
For these low $Z$ target nuclei, nucleon removal due elastic Coulomb
dissociation can be neglected. Rather, the two strong-interaction driven
nucleon removal mechanisms, elastic and inelastic breakup of the projectile,
are the dominant processes \cite{Han03} -- in which the target nucleus remains
in or is excited from its ground state. For reactions on these light target
nuclei the inelastic breakup (or stripping) mechanism drives the larger part
the removal cross section, the fraction due to the elastic breakup mechanism
being between 40\% and 17\% in the cases where the relative yields of these
processes have been distinguished by measurements \cite{Bazin,Wimmer}. The
fractions of these elastic and inelastic removal events predicted by the
eikonal-model calculations, that depend strongly on the nucleon separation
energy, are in excellent agreement with the results of these dedicated
measurements for removals from both weakly-bound \cite{Bazin} and well-bound
\cite{Wimmer} orbitals. Here we discuss only the sum of these two contributions
that, in general, are not distinguished.

Until somewhat recently, measurements have routinely determined only the total
number of bound, mass $A-1$ residues. The cross section measurements presented
here are also inclusive with respect to all bound final states of the reaction
residue. For many of the rare, highly neutron-proton-number asymmetric nuclei
used, these bound final-states spectra are unknown or only partially known.
The model analyses thus take the final-state spins, parities and excitation
energies from shell-model calculations with effective interactions and model-spaces
appropriate to the mass and charge of the projectile. The theoretical inclusive
cross sections are taken as the sum of the calculated partial cross sections to
all of the shell-model states of the residue with predicted excitation energies
consistent with the empirical, if known, or the evaluated lowest particle
emission threshold and its uncertainty; that is, $S_n$ for neutron-rich or
$S_p$ for neutron-deficient residues.

\section{Analysis and Results}

\subsection{Methodology}

We summarize the analysis methodology that is applied to each reaction. The
theoretical partial cross section for removal of a nucleon, from a single-particle
configuration $j^{\pi}$, populating the residue final state $\alpha$ with
excitation energy $E_\alpha^*$, is calculated as \cite{Han03}
\begin{equation}
\sigma_\text{th}(\alpha)=\left(\frac{A}{A-1}\right)^{\!\!N} C^2S(\alpha,j^{\pi})
\,\sigma_\text{sp}(j,S_\alpha^*) \label{eq:xsec} ,
\end{equation}
where $S_\alpha^*=S_{n,p}+E_\alpha^*$ is the separation energy to the final
state $\alpha$ and $S_{n,p}$ is the ground-state to ground-state nucleon
separation energy. $N$, in the $A$-dependent center-of-mass correction to
the shell-model spectroscopic factors, $C^2S(\alpha,j^{\pi})$, is the number
of oscillator quanta associated with the major shell of the removed particle
\cite{Die74}. The single-particle cross section $\sigma_\text{sp}$ is the
sum of the elastic and inelastic breakup contributions to the reaction
\cite{Tos01} calculated assuming the removed-nucleon's single-particle
wave-function (or overlap function) is normalized. The theoretical inclusive
nucleon-removal cross section, $\sigma_\text{th}$, is then the sum of these
partial cross sections to all bound final states of the mass $A-1$ residue.
This cross section is thus the predicted reaction yield resulting from the
single-particle strengths to the low-energy spectrum of the residue. As should
be clear from the above, this inclusive cross section does not probe the values
of individual spectroscopic factors, $C^2S(\alpha,j^{\pi})$. The overall
shell-model strength to bound final states can, however, be compared with
the measured cross sections, $\sigma_\text{exp}$. This is the basis and
physics of the cross sections ratio $R_s = \sigma_\text{exp}/\sigma_\text{th}$.

The asymmetry of the neutron and proton separation energies from the orbitals
near their Fermi surfaces is quantified, in each reaction, by the parameter
$\Delta S$. If only the residue ground state is bound then $\Delta S = S_n-S_p$
for neutron removal and $\Delta S =S_p-S_n$ for proton removal. When there are
several bound shell-model final states the separation energy of the removed
particle in $\Delta S$, above, is replaced by the partial-cross-section weighted
average of these bound-states $S_\alpha^*$. With this convention, the removal of
the most strongly-bound (weakly-bound) nucleons from proton-neutron asymmetric
nuclei have large positive (negative) values of $\Delta S$. The
deduced $R_s$ and $\Delta S$ for the body of data sets are shown in Figure
\ref{fig:one}.

The inputs to the $\sigma_\text{th}$ calculations: the ranges of the optical
potentials and of the nucleon radial overlaps, that dictate the reaction geometry,
must be chosen consistently \cite{Gad08}. Each $\sigma_\text{th}$ calculation
requires realistic: (i) final-states spectra and $C^2 S$, (ii) residue- and
nucleon-target optical potentials and their derived elastic S-matrices, that
localize the reactions spatially, and (iii) removed-nucleon radial wave-function
geometries from the projectile ground state. For (i) we use the best available
shell-model calculations and for (ii) and (iii) we constrain the shapes and
radial size parameters using theoretical, Hartree-Fock (HF) model systematics.
The errors on the calculated $\sigma_\text{sp}$, arising from the precise values
of these sizes (rms radii) of the nucleon radial wave-functions and those of
the residual and target nuclei, with which the cross sections scale essentially
linearly \cite{Gad08}, have been estimated using finite-difference derivatives.
These were summarized in Eq. (3) of Ref. \cite{Terry} and Eq. (5) of Ref.
\cite{32Ar} for removals from weakly-bound and well-bound orbitals, respectively.
The likely errors from this source, typically 10\% (strongly-bound cases) and
5\% (weakly-bound cases) for a $\delta R$ uncertainty of 0.1 fm, are modest
and confirm that the model calculations are robust against plausible changes
to these theoretically-constrained geometric inputs. The variation of the
computed $R_s$ when using different Skyrme forces in the HF calculations used
to constrain these radial sizes were shown to be rather minimal; see for example
Fig. 7 of Ref. \cite{Gad08}. All calculations presented here use the SkX Skyrme
interaction \cite{Bro98}. More complete details of the procedures used, common
to all of the data sets, are detailed in Refs. \cite{Gad08,TG}.

We add that, as most of the earlier data sets analysed, and several of the new
data sets included here, are for beam energies close to 100 MeV/nucleon on a
$^9$Be target, the removed nucleon-target nucleus optical potential and S-matrix
used in the analyses are essentially the same for the majority of the data sets
across the full range of $\Delta S$ values. It follows that details of this
optical potential can play little part in the observed behaviour of $R_s$
with $\Delta S$.

\subsection{Additional Data Sets}

\begin{figure}[h]
\begin{center}
\includegraphics[width=1.07\columnwidth,angle=0,scale=1]{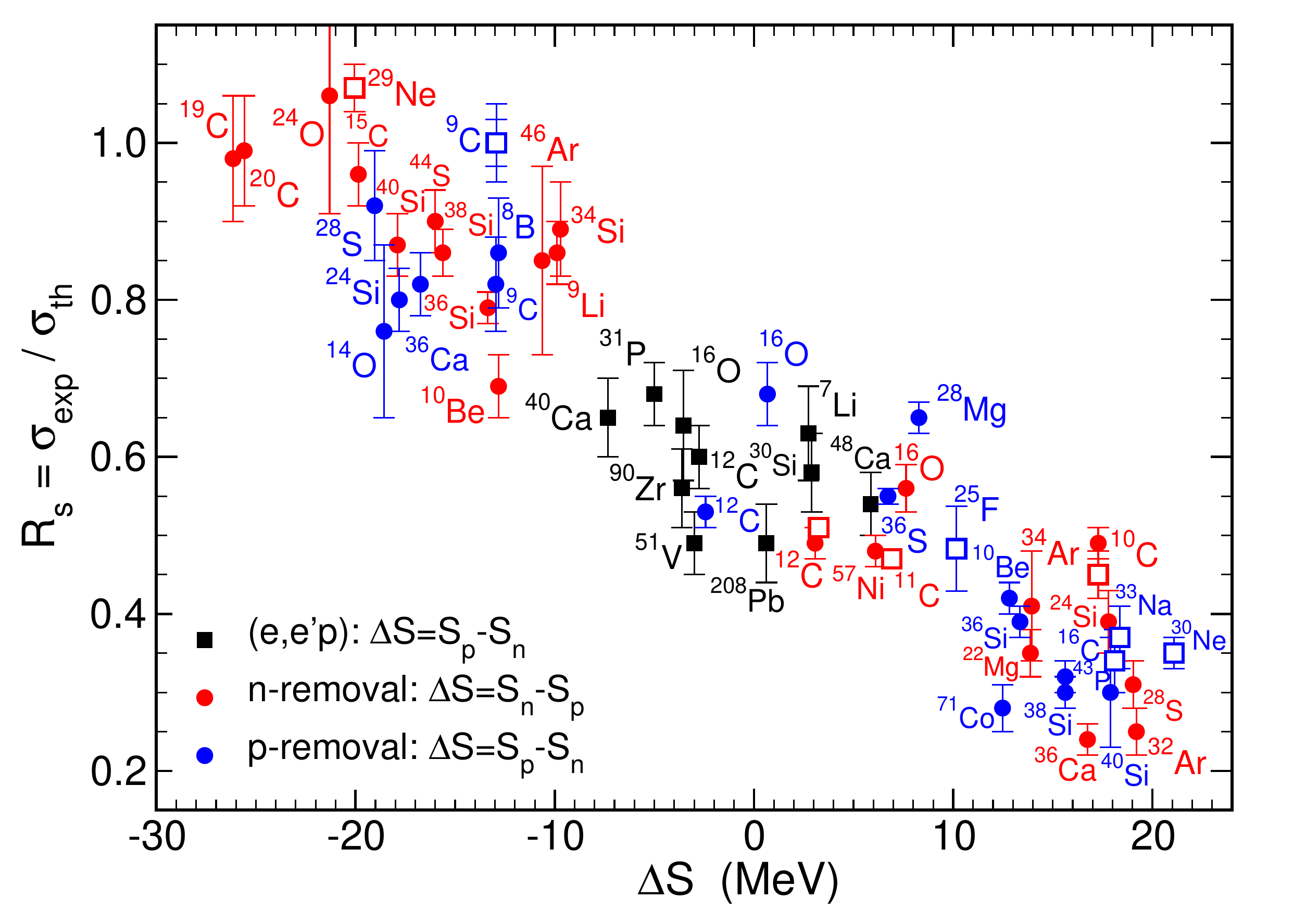}
\end{center}
\caption{Compilation of the computed ratios, $R_s$, of the
experimental and theoretical inclusive one-nucleon removal cross sections
for each of the projectile nuclei indicated. $R_s$ is shown as a function of
the parameter $\Delta S$, used as a measure of the asymmetry of the neutron
and proton Fermi surfaces. The new red (neutron removal) and blue (proton
removal) higher-energy data points are shown as open squares. The solid
(black) squares, deduced from electron-induced proton knockout data, are
identical to those in the original compilation of Ref. \cite{Gad08}.}
\label{fig:one}
\end{figure}

The data compilation in Figure \ref{fig:one} includes all cases reported in
Ref. \cite{TG}. These included a number of higher-energy measurements. For
example, the $^{19,20}$C($-n$) and $^{14}$O($-p$) data points, with large
negative $\Delta S$, were made at 240 and 305 MeV/nucleon, respectively.

The updated figure includes the results of several new reported measurements
in the 80-100 MeV/nucleon range. From left to right in the figure, these added
analyses are for: $^{24}$O($-n$) \cite{Div}, $^{44}$S($-n$) \cite{KW20},
$^{34}$Si($-n$) \cite{Jong}, $^{36}$S($-p$) \cite{Mutsch}, $^{71}$Co($-p$)
\cite{Fe60} and $^{43} $P($-p$) \cite{GadP}. Each of these new measurements
was performed at the National Superconducting Cyclotron Laboratory's Coupled
Cyclotron Facility, located at Michigan State University~\cite{nscl}. All of
these additional systems are seen to be consistent with the established trend
of the collected data.

Of particular interest are the additions, shown as open square symbols,
from higher-energy measurements and model calculations at 220-240 MeV/nucleon
and 1.6 GeV/nucleon. The 220-240 MeV/nucleon additions are, from left to right,
for $^{29}$Ne($-n$) \cite{Kobay}, $^{25}$F($-p$) \cite{Yoshi}, $^{16}$C($
-p$) \cite{Zhao}, $^{30}$Ne($-p$) \cite{Liu}, and $^{33}$Na($-p$) \cite{Murray}.
For the $^{25}$F($-p$) measurement, made on a carbon target at 218 MeV/nucleon,
the data point is computed using the inclusive cross-section reported informally
in Ref.\ \cite{Yoshi}. The shell-model calculation in that case used the universal
$sd$-shell {\sc usdb} effective interaction \cite{usdb}, the only significant
transition being $d_{5/2}$ proton removal populating the $^{24}$O ground-state
with a spectroscopic factor of 1.01 \cite{Tang}. This gives an $R_s$ value of
0.48(5) with $\Delta S=10.17$ MeV.

All measurements, except that for $^{16}$C($-p$), were performed at the
Radioactive Isotope Beam Factory (RIBF) at RIKEN, Japan. The $^{16}$C measurement
was made at the External Target Facility, Institute of Modern Physics, Lanzhou,
China. All data sets have been reanalysed with the earlier-stated methodology,
so the deduced $R_s$ may differ in detail where published values used different
inputs. Specifically, the $^{16}$C($-p$) data point in Fig.\ 1, deduced from
the cross section reported in Ref. \cite{Zhao}, is $R_s=0.34(4)$ with a
$\Delta S$ value of 18.1 MeV.

The added data points at 1.6 GeV/nucleon are for the carbon isotopes:
$^{9}$C($-p$), $^{12}$C($-n$), $^{11}$C($-n$) and $^{10}$C($-n$), measured
at the GSI Helmholtzzentrum f\"ur Schwerionenforschung, Darmstadt, Germany.
The analyses of these measurements are detailed in Ref. \cite{Schlemm}. For
the new $^{9}$C($-p$) and $^{10}$C($-n$) reaction data, where there were
also earlier, lower-energy analyses \cite{Gri12} at 100 and 120 MeV/nucleon,
respectively, the agreement of the deduced $R_s$ at the two energies is
excellent.

\begin{figure}[t]
\begin{center}
\includegraphics[width=1.04\columnwidth,angle=0,scale=1]{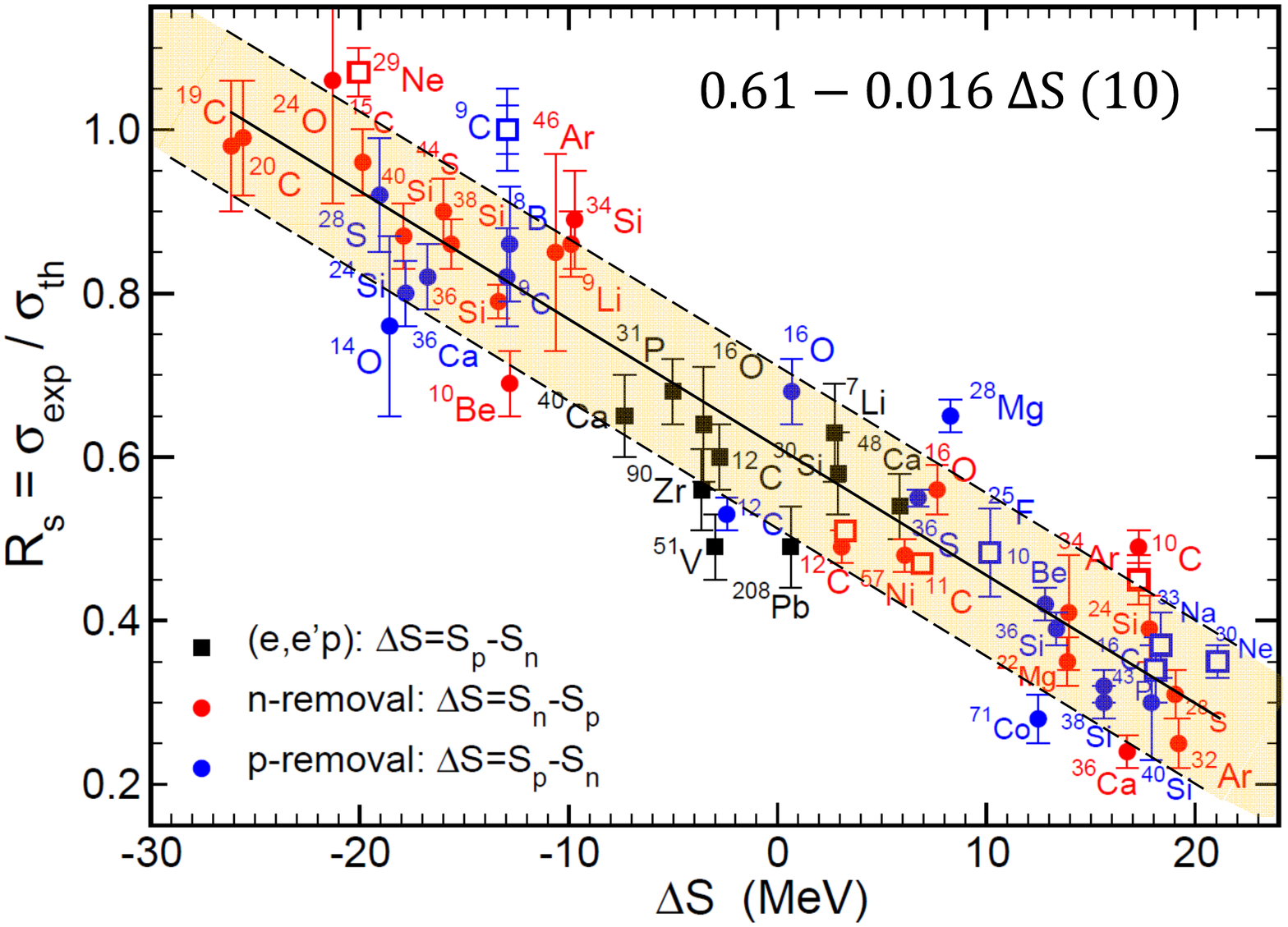}
\end{center}
\caption{As for Fig. 1. The trend line given in Eq. \ref{eq2}, as reported
in Ref. \cite{Mutsch}, and a band of half-width 0.1 (shaded region)
have been superimposed, which summarize the totality of collected data
points.} \label{fig:two}
\end{figure}

Each of these independent data analyses, for projectiles across the nuclear
chart and spanning the full extent of nucleon separation energies (and
$\Delta S$), are found to fall within the band of scatter of values of the
earlier published systematics \cite{TG}. We note that, as was stated in
Ref. \cite{Mutsch}, and as presented graphically in Fig. \ref{fig:two},
a linear representation of the entire collection of data in Fig. \ref{fig:one}
is provided by the trend-line
\begin{equation} \label{eq2}
R_s = 0.61-0.016\, \Delta S
\end{equation}
together with a scatter in the deduced $R_s$ values, from the different
regions of the nuclear chart, of order 0.1, independent of the $\Delta S$
value. This scatter of deduced values is not surprising given the very
diverse set of projectile nuclei involved and the distinct shell-model
spaces and effective interactions involved in the analyses of each of
the different mass and charge regions.

It is important to stress that the presented $R_s$ systematics provide an
overall, sum-rule-like, inclusive measure of the theoretically-predicted
strength of transitions to bound final states, and do not provide information
on any individual transition or spectroscopic factor. In particular, $R_s$
is not an overall scaling factor to be applied to each contributing theoretical
partial cross-section. The relationship between the model and measured partial
cross sections is expected to be far more complex. For example, in the recent
$^{34}$Si($-n$) reaction analysis included above \cite{Jong}, the measured and
model partial cross sections to $3/2^+$ and $1/2^+$ $^{33}$Si final states are
in broad agreement. On the other hand, the model calculations predicted a number
of strongly-populated $5/2^+$ final states, that enhance the theoretical inclusive
cross section, wheras these $5/2^+$ states were found to be more weakly populated
in the measurements. So, to better-interpret the deduced $R_s$ value in each case
requires more final-states-exclusive studies, as are now possible more routinely
given the ongoing improvements in both beam intensities and detection capabilities.

\subsection{Discussion}

For the majority of spectroscopic applications, the details of these absolute
cross section comparisons, as quantified by $R_s$, are secondary. As in
essentially all direct-reaction applications, principally, it is comparisons of
the measured and theoretical relative partial cross-sections to each final state
and, in intermediate-energy removal reactions, the shapes of their residue momentum
distributions, that allow both the dominant transitions and single-particle
configurations of the valence nucleons in the projectile ground state to be
identified. The absolute inclusive cross section comparisons, shown in Figs. 1
and 2, pose additional and challenging questions of both the reaction dynamics
treatment and the nuclear structure model inputs. The following discussion
summarizes aspects of ongoing investigations relevant to these open questions.

The physical origins of the presented systematic behaviour, with a reduction
of $R_s$ with increasing $\Delta S$, obtained consistently from the eikonal-
plus sudden-approximation reaction dynamics and shell-model nuclear structure,
remain unresolved. Testing for inadequacies in the reaction dynamics treatment,
e.g. of the sudden approximation, was a motivation of the present work -- using
higher-energy data sets. Concerns regarding the shell-model nuclear-structure
input, in particular the role and treatment of correlations in the many-body
wave functions, are also relevant. The localization of the reaction, discussed
earlier, means that absolute cross sections are particularly sensitive to the
wave functions (overlap functions) of the removed nucleons near the nuclear
surface, where the importance of neutron-proton correlations is an open question.
Specifically, in reactions with large positive $\Delta S$ and at radii near the
nuclear surface, the removed, minority nucleon species occupying orbitals near
their energetically well-bound Fermi surface are embedded, spatially, within
nuclear matter that is dominated by the other, majority species. We return to
this discussion below.

Given these open questions, other nucleon-removal mechanisms, direct-reaction
models and experimental data have also begun to be used to probe these absolute
cross-section effects. Ref. \cite{Aumetal} presents a comprehensive recent review
of these studies which, to date, involve a more limited set of projectile species
than used here. These alternative reactions involve different spatial localizations
and radial sensitivities, and bring their own approximation schemes, model inputs
and uncertainties. For example, measured and theoretical cross-section ratios
have been presented using analyses of single-nucleon transfer reactions
\cite{Lee:2011,Flavigny:2013,Kay:2013}, analysed using conventional distorted-waves
Born approximation-like techniques, and inverse kinematics $(p,2p)$ and $(p,pn)$
knockout processes on a proton target, analysed using quasi-free scattering
\cite{Atar,Kawase} and coupled-channels approaches \cite{Ramos}. In the case
of the transfer reaction analyses, experimental (beam-intensity and target-thickness)
requirements have limited the range of $\Delta S$ values accessible, while the
ratios presented are generally from selected exclusive (e.g. ground-state) cross
sections and not from the bound-final state inclusive yields. The $(p,2p)$ and
$(p,pn)$ knockout analyses cover a wider range of $\Delta S$ values but have
been concentrated on the oxygen and carbon isotopes, with very few cases in
common with the compilation in Figure 1. An exception is an inverse
kinematics $(p,2p)$ measurement of the $^{25}$F($-p$) reaction \cite{Tang},
that deduced an empirical spectroscopic factor of 0.36(13) to the $^{24}$O
ground state, the only bound final state. The corresponding new measurement on
a C target was included above. Compared to shell-model values, the spectroscopic
factor from this $(p,2p)$ measurement is more heavily suppressed than the value
from the higher-precision C target data, from which the deduced empirical
spectroscopic factor, $\sigma_\text{exp}/\sigma_\text{sp}$, is 0.53(6). A
programme of precision measurements, on a common set of projectile and residue
nuclei, using these alternative removal mechanisms -- with their different spatial
sensitivities to the wave functions of the removed nucleons -- would be of
considerable value in making direct comparisons.

Overall, and based on their more-limited reaction data sets, the ratios of the
measured and reaction model cross sections from the transfer, $(p,2p)$ and
$(p,pn)$ reaction analyses are more constant as a function of $\Delta S$ than is
shown in Figures 1 and 2. In contrast, a recent and completely different reaction
methodology -- the intra-nuclear-cascade (INC) model \cite{DiazCortes:2020} --
applied to inclusive cross sections from fast neutron- and proton-removal reactions
from medium-mass neutron-rich nuclei on a beryllium target at energies near 1
GeV/nucleon, deduced highly-asymmetric $\sigma_\text{exp}/\sigma_\text{th}$ ratios
between the neutron- and proton-removal channels. There, as in the present work,
the calculated cross-sections for the well-bound, large positive $\Delta S$
proton-removal cases significantly exceeded the measured values. The conclusion
drawn from that study was that the absence of a realistic treatment of short-range
correlations in the INC-model calculations was principally responsible for this
outcome. As that more-macroscopic model approach also uses only very limited nuclear
structure input -- and not the detailed shell-model final states and spectroscopic
information used in the model of this work -- the ratios of cross-sections deduced
in Ref. \cite{DiazCortes:2020} are not the same as the $R_s$ defined here.

Regarding these shell-model theoretical inputs, it is reasonable to expect that
the actual physical spectroscopic strengths will be somewhat suppressed compared
to those computed in the highly-truncated model-spaces of practical shell-model
calculations. This effect has not yet been fully quantified. Relevant, however,
are the extensive recent nucleon-nucleon (NN) short-range correlation (SRC) studies
probed using high-energy electron beams, e.g. \cite{Hen2,Duer1,Duer2}. These data
show conclusively that, for higher nucleon momenta, SRC effects are dominated by
neutron-proton pairs and that, in nuclei with a neutron excess, the minority
proton species have a significantly greater probability to be found with momenta
above their Fermi momentum than do the neutrons. This was interpreted as due to
the action of the NN tensor force. A phenomenological approach to see the effect
of these correlations on calculated electron- and proton-induced knockout cross
sections from stable and asymmetric nuclei was presented in Ref. \cite{Pasch}.
A conclusion was that the increased proton high-momentum components may, under
certain assumptions, result in reduced occupancies and spectroscopic factors for
proton orbitals near their Fermi-surface. Our expectation is that the action of
these neutron-proton SRCs, between the minority valence protons and the excess
of neutrons, that occupy less-well-bound but spatially-overlapping orbitals, will
elevate some fraction of the valence protons to otherwise empty orbitals above
their Fermi-surface. Such excited, normally-unoccupied proton orbitals are not
conventionally included in the spaces of shell-model calculations, as used in
this work. The shell-model occupancies of the valence nucleons, confined to
orbitals in the restricted model-space, are thus expected to be somewhat enhanced.
This effect needs further and quantitative investigation.

The major part of these alternative efforts, and the present analysis, have
concentrated on the reduced $R_s$ values at large positive $\Delta S$, those
most likely to be impacted by any failings of the sudden approximation to the
dynamics. There remains very limited discussion of the deduced larger $R_s$ in
systems with large negative $\Delta S$ -- the removal of very weakly-bound and
halo-like nucleons. These $R_s$, with values near unity, are an equally strong
feature of the data and model systematics and, based on the discussions above,
signal that the correlations experienced by these spatially-extended valence
nucleons -- beyond those included approximately through the shell-model effective
interactions -- may be significantly reduced. However, since there is no explicit
treatment of the continuum in the shell-model calculations used here, these
questions cannot be addressed quantitatively within the presented model.
We have shown, however, that these systematics are a robust feature of the
eikonal- plus shell-model analyses and that these are reinforced by the
additional, newly-available measurements, that include data at significantly
higher beam energies.

\section{Conclusion}

We have reported comparisons of measured and calculated inclusive single nucleon
removal cross sections for additional reactions, including many at significantly
higher beam energies. For these higher-energy data, the collision time is correspondingly
shorter and the validity of the sudden approximation in the collision dynamics
is thus enhanced. Analyses of these new measurements, that include energies from
around 100 MeV/nucleon up to 1.6 GeV/nucleon, agree with the earlier-noted trends
of the ratio of the measured and theoretical cross sections, $R_s = \sigma_\text{
exp}/\sigma_\text{th}$, with the neutron-proton separation energy asymmetry
parameter $\Delta S$. The results from these new data sets, that span the full
range of $\Delta S$ values, show no significant deviations from the earlier
data analyses or indicate any breakdown of the sudden-approximation in the analyses
of these and the numerous other measurements in the energy range of 80--120 MeV/nucleon.

\begin{acknowledgments}
The authors thank Profs. Joachim Enders (TU Darmstadt) and Takashi Nakamura
(Tokyo Institute of Technology) for information on and citations to their
Groups' recently-analysed data sets. This work was supported by the United
Kingdom Science and Technology Facilities Council (STFC) under Grant No.
ST/F005314/1 and by the U.S. Department of Energy, Office of Science, Office
of Nuclear Physics, under Grant No. DE-SC0020451.
\end{acknowledgments}

\bibliographystyle{apsrev}

\end{document}